# A novel method for measurement of the refractive indices of transparent solid media using laser interferometry

Arnab Pal[1], Kriti R. Sahu[2], Pradipta Panchadhyayee[1*] and Debapriyo Syam[3]

[1]Department of Physics (UG & PG), P.K. College, Contai, PurbaMedinipur 721404, West Bengal, India, arnab95pal@gmail.com

[2]Department of Physics, Egra SSB College, Egra, PurbaMedinipur 721429, West Bengal, India, kriti.sahu91@gmail.com

[3]CAPSS, Bose Institute, Salt Lake City, Kolkata 700091, West Bengal, India, syam.debapriyo@gmail.com.

*Corresponding author: ppcontai@gmail.com

## ABSTRACT

A novel method is proposed to measure the refractive indices (RIs) of the materials of different transparent solid state media. To exploit the advantage of non-contact measurement laser beam interferometry is used as an effective technique for this purpose. The RIs of materials are also experimentally determined with the aid of another laser based simple method. The derivations of the working formulae for both the methods are presented. The experimental values of RI of any glass sample found by the different methods are consistent with each other and fall within the range of known values of RI for glass. Both types of experiment can be set up rather easily in an undergraduate laboratory. They can supplement other methods of finding refractive index like, for example, methods based on the use of a spectrometer.

## Introduction:

Refractive index is a number which governs how light changes its direction of propagation as it enters one material medium from another. This phenomenon is known as refraction and the angles of incidence and refraction of light, referred to the normal to the interface of the two media at the point of incidence, are related by Snell's law. Refractive index (RI) depends on the color (or wavelength $\lambda$) of light [1]. Tables of values of refractive index for various media and wavelengths of light, with respect to vacuum, are readily available [2, 3]. Refractive index of a material can be measured by many methods; for example by using a spectrometer in conjunction with a prism made of the experimental substance. An important class of methods of measuring RI involves the formation of interference patterns. Interferometric measurements are concerned with the study of separation between bright fringes (or dark fringes) resulting from the superposition of light waves, originating from a single source and propagating along paths of different optical lengths (refractive index multiplied by geometric path length). Researchers have exploited the scope of availing nearly monochromatic light from the laser sources to enhance precision in measuring refractive index by applying the interferometric techniques.

In this article we present two simple laser-based experimental arrangements to determine the values of RIs of transparent materials. Both experiments can be set up without much trouble and one of the experimental methods leads to spectacular circular fringes and for that reason alone, if not also for the determination of RI, that experiment can and should be set up.

## Theoretical Background:

Researchers have focused on the use of laser [4, 5] for the measurement of RI, instead of conventional light sources, due to the exceptionally high coherence properties of laser. The interference patterns exhibit extreme sharpness leading to much better experimental accuracy. In the present work, as mentioned above, we present two simple experimental arrangements to determine the values of RIs of transparent materials. In the first method we sprinkle particles of chalk powder over the free surface (top of glass or plastic covering slab) of a mirror. A laser beam is allowed to strike some powder particles located on the free (top) surface of the transparent slab. The rays get scattered in all directions. One of the scattered rays (i.e. a ray resulting from the scattering of an incident ray), on hitting the top surface of the slab, splits into a reflected ray and a refracted ray. The reflected ray leaves the top surface at an angle $\alpha$ (see Fig.

1) to the normal. The refracted ray enters the slab at an angle $\beta$ to the normal and after reflection at the back surface and subsequent refraction at the front surface, produces an emergent ray that makes the same angle α with the normal to the front surface. The first reflected ray and the emergent ray interfere with each other at a large distance (very large in comparison with the diameter of the laser beam) from the mirror and, together with other similar pairs of rays, form a sharp interference pattern. We can easily take snapshots of the pattern by a high resolution camera and analyze the intensity distribution by the 'ImageJ' software [6]. Subsequently the RI of the transparent material can be found using the relevant formula which is derived below.

In the first method we note that the pattern is the natural consequence of interference of the first scattered ray (or wave) and the refracted ray (or wave). As is seen in Fig. 1, the phase difference between them is (a) $2\mu d\cos\beta\times2\pi/\lambda$, or (b) $2\mu d\cos\beta\times2\pi/\lambda-\pi$, depending on how phase changes on reflection, where $\mu$ is the refractive index of the slab. Other scattered rays that enter the transparent slab at angles different from $\beta$ interfere with the first scattered ray close to the mirror; they are not responsible for the interference pattern produced at a large distance.

Here, for case (a), if $2\mu d\cos\beta = n\lambda$, then constructive interference takes place and a bright ring is produced [4]. On the other hand, if $2\mu d\cos\beta = (2n+1)\lambda/2$, then a dark ring results. (Note that for $\beta\approx 0°$, $n\sim10^4$, assuming d ~ 2 mm). If $\delta\beta$ be the change of $\beta$ between the $n$th and $(n+1)$th dark rings we can write the relation:

$$2\mu d\times\delta(\cos\beta) = 2\mu d\times\sin\beta\delta\beta = \lambda \quad (1)$$

Also,

$$sin\alpha = \mu sin\beta \quad (2)$$

As, in practical situations, $\beta$ is a very small angle, $\delta\beta = \delta\alpha/\mu$. Therefore

$$\mu = 2d\ (\sin\alpha\delta\alpha/\lambda).$$

The value of $\mu$ can thus be calculated by measuring $d$, $\sin\alpha$ and $\delta\alpha$. For a typical dark ring,

$$\sin\alpha \approx \tan\alpha\ =\ R/D$$

where the screen, on which the rings are observed, lies at a distance $D$ from the front surface of the mirror (see Fig. 3) and $R$, is the radius of that ring. Thus we finally obtain

$$\mu \approx 2d\ (R\delta R/\lambda D^2) \quad (3)$$

It is worth mentioning that order of the fringe ($n$) decreases while the radius of the circle produced on the screen increases with $\beta$. Also, because of the faster variation of $\cos\beta$ with increasing $\beta$, the separation between fringes i.e. the difference of radii decreases with the decrease in $n$ (as observed). This is mathematically established below:

First, we have, using the previous equations

$$\frac{dR}{d\beta} = \left(\frac{dR}{d\alpha}\right)\frac{d\alpha}{d\beta} = D\left(\frac{\mu cos\beta}{cos^3\alpha}\right) \qquad (4)$$

Now the typical value of $n$ is of the order of $10^4$. Let $\delta R$ represent the change of $R$ for a small variation $(\delta n)$of the order number of a fringe. Since $n \gg \delta n$, we may write

$$\frac{\delta R}{\delta n} \approx \frac{dR}{dn} = \frac{dR}{d\beta}\left(\frac{d\beta}{dn}\right) = D\left(\frac{\mu cos\beta}{cos^3\alpha}\right)\frac{(-\lambda)}{2\mu d sin\beta}.$$

Rewriting

$$\frac{dR}{dn} = -D\left(\frac{\mu cos\beta}{cos^3\alpha}\right)\frac{\lambda}{2d sin\alpha}$$

$$= -\frac{\mu\left(R^2+D^2\right)^2\lambda}{2dD^2R}\left[1 - \frac{R^2}{(R^2+D^2)\mu^2}\right]^{1/2} \qquad (5)$$

since$sin\alpha = \frac{R}{\sqrt{R^2+D^2}}$, $cos\alpha = \frac{D}{\sqrt{R^2+D^2}}$ and $cos\beta = \left[1 - \frac{R^2}{(R^2+D^2)\mu^2}\right]^{1/2}$.

Note that $\left|\frac{dR}{dn}\right|$ increases (decreases) with $n$ i.e. towards smaller (larger) value of$\beta$. (Since $R \ll D$, in practical situations, $\left|\frac{dR}{dn}\right| \sim \frac{1}{R}$). This means that separation between fringes (i. e. their radii) decreases with the increase of $R$. This is in accordance with experimental results. <u>To improve clarity, we represent the radius ($R$) of the ring of order $n$ by $R_n$</u>. Then

$$\frac{dR_n}{dn} = \frac{-\mu\left({R_n}^2+D^2\right)^2\lambda}{2dD^2R_n}\left[1 - \frac{{R_n}^2}{({R_n}^2+D^2)\mu^2}\right]^{1/2} \qquad (6)$$

Using equation (6) we can calculate the value of the refractive index of an unknown transparent material:

$$\mu = \frac{R_n}{\sqrt{({R_n}^2+D^2)}}\left[1 + \frac{\left(\frac{dR_n}{dn}\right)^2 4d^2D^4}{\left({R_n}^2+D^2\right)^3\lambda^2}\right]^{1/2} \qquad (7)$$

Note also that the **observed pattern is not an Airy pattern** (reflected by the mirror). The difference of radii of adjacent rings, in an Airy pattern, is almost a constant contrary to our observation. Moreover, in the Airy pattern, the sizes of the rings depend on the size of the circular aperture through which the laser beam passes before striking the mirror. That is not the case here. Again, chalk-powder, an essential element in our experiment, can play no role in the formation of an Airy pattern.

The pattern being reported is also different from the 'Quetelet fringes' [7]. Quetelet fringes are produced by dust particles which have settled down on a mirror, but they are *localized fringes* i.e. they form close to the surface of the mirror, unlike the fringes observed by us.

In the second method, a direct procedure for measuring the RI of the transparent material is proposed that uses the arrangement shown in Fig. 2. We show how, by simply using a travelling microscope and a screw gauge, the RI of the material of a transparent slab (e.g. glass top of a mirror) for the light in a laser beam can be measured. This value of RI is then compared with the value of RI obtained by the first method.

Here, the same laser beam is allowed to fall on the glass top of the mirror at an angle "$i$". After refraction (at an angle "$r$"), and further reflection and refraction, a part of the beam is emitted from the top surface and it follows the path shown in Fig. 2(a).

It is obvious from Fig. 2 (a) that

$$y = 2d\, tan(r) \tag{8}$$

Therefore,

$$\sin(r) = sin\left(tan^{-1}\left(\frac{y}{2d}\right)\right) \tag{9}$$

Thickness, *d*, can be found with a screw gauge with least count (l.c.) = 0.01 mm, and separation '$y$' by using a travelling microscope (also with l.c. = 0.01 mm). When the spots overlap, as shown in Fig. 2(b), $y$ can still be found by virtue of the difference in brightness of the spots produced on the top surface. The angle, $i$, can be found with the help of a protractor or by an ingenious method as described later. The value of $\mu$ can then be calculated using Snell's law:

$$\mu = \frac{\sin\,(i)}{\sin\,(r)} \tag{10}$$

**Experimental details:**

**First Method:**

We took clean plane thin mirrors (sample I – III) made of glass (ordinary flint)with different thickness (sample-I: 12.33×11.11×1.77 $mm^3$, sample-II: 11.61×12.52×3.25 $mm^3$ and sample-III: 18.65×17.75×4.82 $mm^3$), and a red transparent plastic sheet with reflecting coating at the back surface (sample-IV: 25.65×25.75×0.8 $mm^3$); the RIs of the transparent covers of mirrors were to be determined. Later on, for convenience, we refer to sample IV as the red plastic mirror.

For the experiment each of them was placed on a stand in front of a laser source emitting red light of wavelength 660 nm. A screen with a central hole of diameter ≈ 2.5 mm was placed between the source and glass/plastic mirror. Source, screen and sample were set up on an optical bench of length ≈ 150 cm. For sample-I (glass mirror of thickness $d$ = 1.77 mm, $D$ was 87.3 cm), for sample-II (glass mirror with $d$ = 3.25 mm, $D$ = 112.3 cm), for sample-III (glass mirror with $d$ = 4.82 mm, $D$ = 117.2 cm) and for sample-IV (red plastic mirror with $d$ = 0.80 mm, $D$ = 107.4 cm). Top surface of the glass/plastic mirror was covered by fine chalk dust. The beam coming out from the laser source passed through the central hole of the screen and fell on the mirror perpendicularly. See Fig. 4(a).The presence of dust particles caused the change in the direction of the beam by scattering. The scattered light after bouncing back from the mirror produced, through interference, alternate red and dark circular rings on the screen. The patterns of dark and red rings have been captured by a digital camera (Canon EOS 750D, focal length 55mm), which was placed behind the mirror.

The images of rings have been analyzed using the 'ImageJ' software. The position of the source of the laser beam and the sample position were kept fixed and patterns [Fig. 4(b)] were photographed for different positions of the screen.

**Second Method:**

The red laser beam fell on a particular sample at an angle. A semi-transparent screen (like, butter paper) was placed in such a way that the incident light beam and the reflected light could pass through it. The distance ($z$) between two laser beam light spots and the perpendicular distance between the semi-transparent screen and sample position ($h$) were measured to calculate the angle of incidence, $i = tan^{-1}(z/2h)$ (shown in Fig. 5) and then the RI of the material of mirror could be found out.

**Results and Discussion:**

For sample I we present part (a) of Fig. 6 to exhibit the graphical profile of intensity corresponding to the interference pattern obtained. Note that the peaks and troughs in the intensity profile correspond to the bright and dark rings respectively. Regarding the variation of the radius of a dark ring ($R_n$(exp)) as a function of the shift of order-number ($\delta n$),where $\delta n$= (order no. of the central spot - order no. of a fringe), as are seen in the (b) part of Fig. 6 and (a) parts of Figs. 7-9, we find that the nature of variation is the same for all the graphs. From the intensity profiles like Fig. 6(a) the values of $R_n$(exp) are measured using pixel size : length relationship in ImageJ [6] via a comparison with a line of known length drawn on the screen. In the above mentioned figures the black curves are drawn by the B-spline fitting of data. Again, in part (c) of Fig. 6 and all the (b) parts of Figs. 7-9, the experimental values of $dR_n/dn$, using Eq. (3) and Eq. (6), are shown against $R_n$. The order numbers of the central spot in the samples I – IV are estimated to be 8360, 15250, 22740, and 3920, respectively, by putting the values of refractive indices (which are obtained through the best fit of Eq. (6) to the $dR_n/dn$ graphs) in the equation $2\mu d = (2n_{central} + 1)\ \lambda/2$.

As mentioned above, experimental values of RIs ($\mu$) of the samples have been deduced from the data using the expression (3) and again with the help of expression (6). Specifically, values of $R$ corresponding to pairs of adjacent data points were used to calculate a series of values of $\mu$ using Eq. (3). The average, $\mu_{1(3)}$, of these $\mu$ values for a sample is quoted in Table 1. Then, varying $\mu$ about $\mu_{1(3)}$, a value of $\mu$ was found that resulted in the best (least-square) fit of Eq. (6) to the data on $dR_n/dn$. This new value of $\mu$ has been referred to as $\mu_{1(6)}$ in Table 1. These RI values are mentioned in all the Figs. 6-9. Some differences exist between the values of ($dR_n/dn$) obtained by using the two alternative methods for all the samples except Sample II. A remarkable agreement is found for Sample II.

A graphical presentation of experimental values of RI of the four samples is made in Fig.10 for better appreciation of the agreement of the values of $\mu_{1(3)}$ and $\mu_{1(6)}$.

**Table 1:** Comprehensive chart of the values of refractive index for glass mirrors with different thickness (Sample-I, Sample-II and Sample-III) and the red plastic mirror (Sample-IV).

| Sample name | First Method | | Second Method |
|---|---|---|---|
| | $\mu_{1(3)}$ | $\mu_{1(6)}$ | $\mu_2$ |
| Sample-I | 1.57 ± 0.01 | 1.56 | - |
| Sample-II | 1.54 ± 0.02 | 1.55 | 1.56 ± 0.04 |
| Sample-III | 1.55 ± 0.03 | 1.56 | 1.56 ± 0.01 |
| Sample-IV | 1.64 ± 0.04 | 1.62 | - |

The experimental values of RI of the three glass samples are in fairly close agreement as is evident from Table 1, and compatible with known refractive indices of common varieties of Crown and Flint Glass $[1.48 \le \mu_{\text{Crown}} \le 1.61, 1.53 \le \mu_{\text{Flint}} \le 1.65]$ [2,3]. The experimental RI values of the samples I and IV by the second method are not given in the table because the '$y$' values were too small to be measured for small thickness of the samples.

**Conclusion:**

A new method, involving laser beam interferometry, has been used to measure the RIs of the materials of different transparent solid state media. The RIs of some of the samples have also been determined by another laser based simple method. The underlying theories of both the experimental methods have been presented in the article. The experimental values of RIs of the samples are compatible with those found in standard literature. The methods are so simple that the experiment can be set up in most undergraduate Physics laboratories and the circular fringes obtained in the first method are truly spectacular! That last feature prompts us to recommend the setting up of the experiment at least for the purpose of demonstration.(We understand that most teachers would still prefer to stick to one of the standard methods as their first-choice method.) One word of caution is in order: Our methods use a laser beam which is potentially dangerous and due care should be taken while setting up the experiment and making measurements so that the eyes are not harmed.

**Acknowledgements:** We gratefully acknowledge the Research Centre in Natural Sciences of P. K. College, Contai and the authority of Egra S. S. B. College for providing Laboratory support.

**Figure Captions:**

Fig. 1: Schematic diagram to determine the RI of mirror materials (First Method).

Fig. 2: Schematic diagram to determine the RI of mirror materials (Second Method).

Fig. 3: Schematic diagram of reflected (refracted) ray and dark or bright circular ring.

Fig. 4: Experimental setup to determine the RI of mirror materials (First Method).

Fig. 5: Experimental setup to determine the RI of mirror materials (Second Method).

Fig. 6: (a) The interference pattern (Inset) and plot of intensity variation for the sample-I (glass mirror with thickness $d$ = 1.77 mm and $D$ = 87.28 cm), (b) variation of the radius of dark ring ($R_n$) with the shift in order number ($\delta n$) and (c) variation of ($dR_n/dn$) with radius of dark ring ($R_n$). Subscript (3) and (6) indicate Eq. (3) and Eq. (6), respectively.

Fig. 7: (a) Variation of the radius of dark ring ($R_n$) with the shift in order number ($\delta n$) and (b) Variation of ($dR_n/dn$) with radius of dark ring ($R_n$) for the sample-II (see text). Subscript (3) and (6) indicate Eq. (3) and Eq. (6), respectively.

Fig. 8: (a) Variation of the radius of dark ring ($R_n$) with the shift in order number ($\delta n$) and (b) Variation of ($dR_n/dn$) with radius of dark ring ($R_n$) for the sample-III (see text). Subscript (3) and (6) indicate Eq. (3) and Eq. (6), respectively.

Fig. 9: (a) Variation of the radius of dark ring ($R_n$) with the shift in order number ($\delta n$) and (b) Variation of ($dR_n/dn$) with radius of dark ring ($R_n$) for the sample-IV (see text). Subscript (3) and (6) indicate Eq. (3) and Eq. (6), respectively.

Fig. 10: Graphical presentations of experimental values ($\mu_{1(3)}$: black color solid circle; and $\mu_{1(6)}$: half filled blue color diamond shapes) of refractive index of three glass mirrors of different thickness (Sample-I, Sample-II and Sample-III) and red plastic mirror (Sample-IV). Vertical line indicates the error bar for the experimental value $\mu_{1(3)}$.

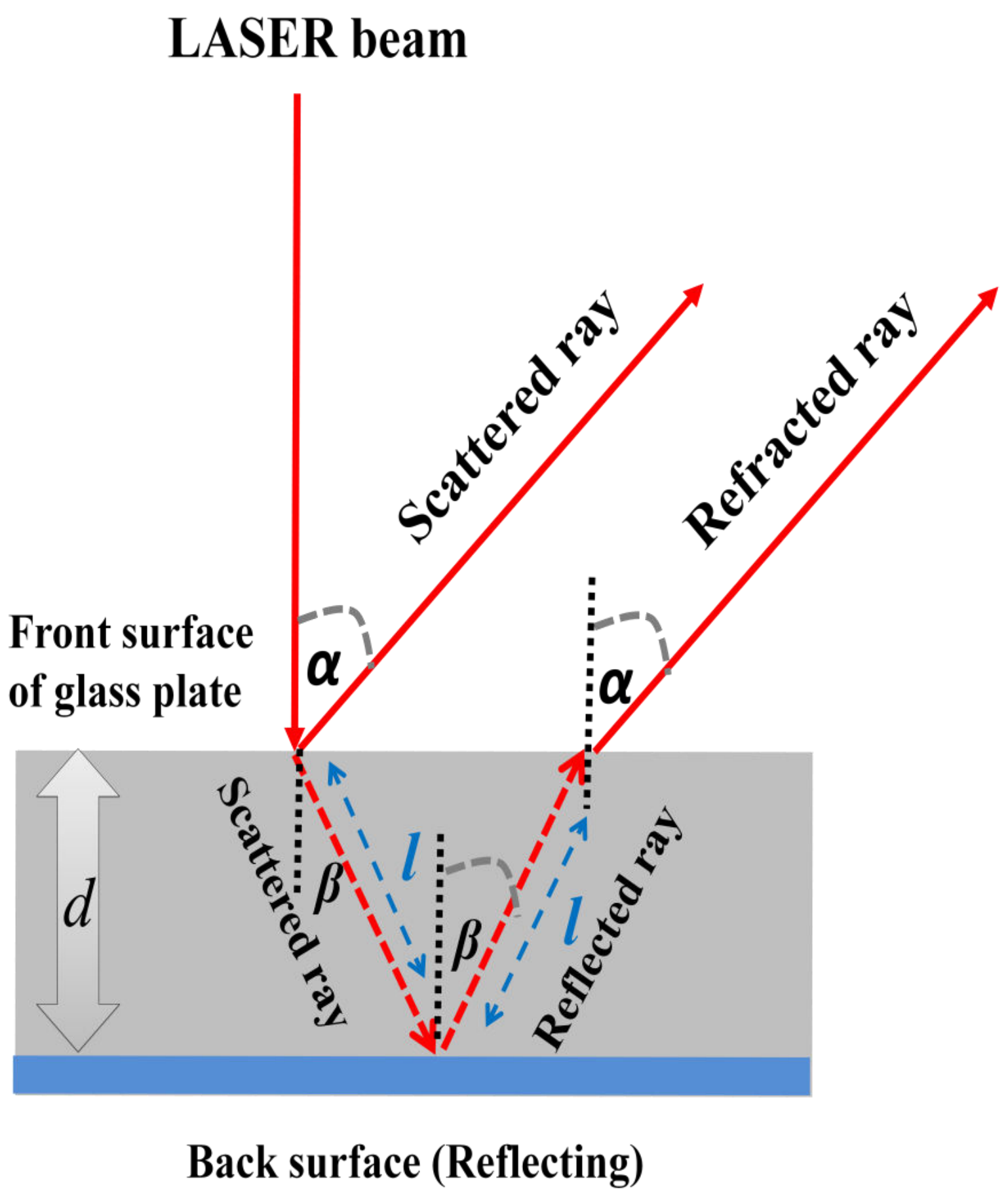

**Fig. 1**

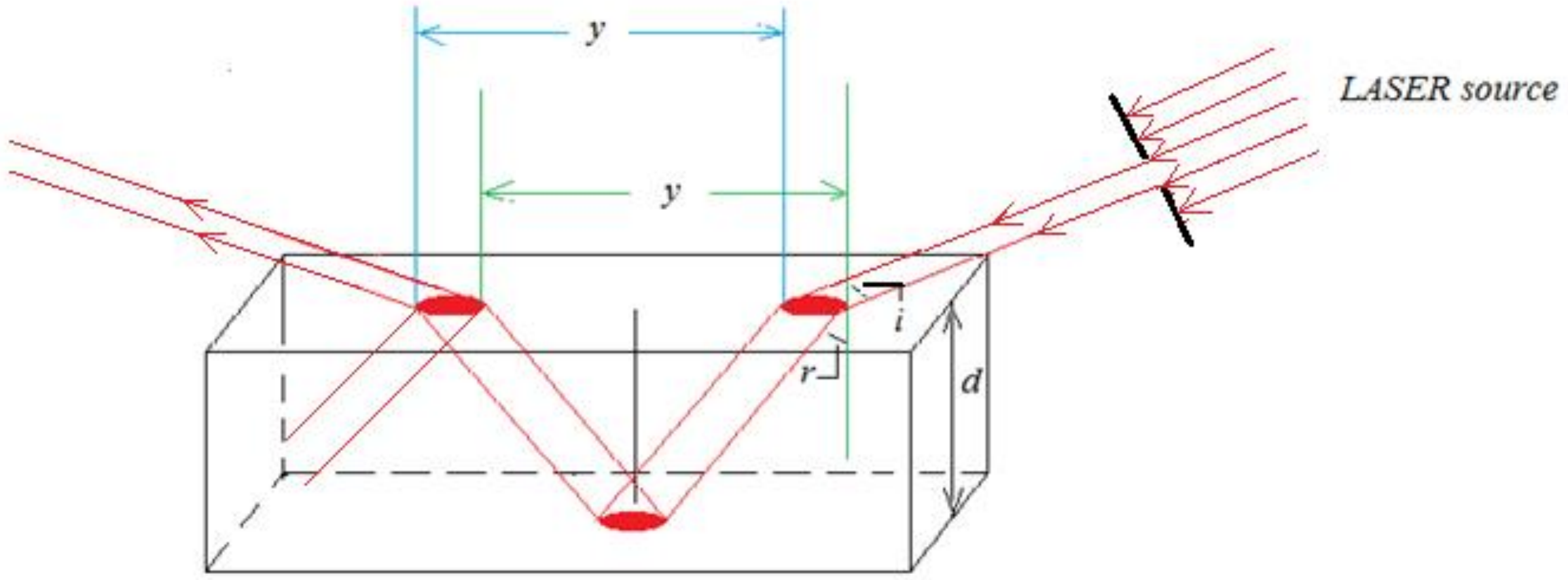

(a)

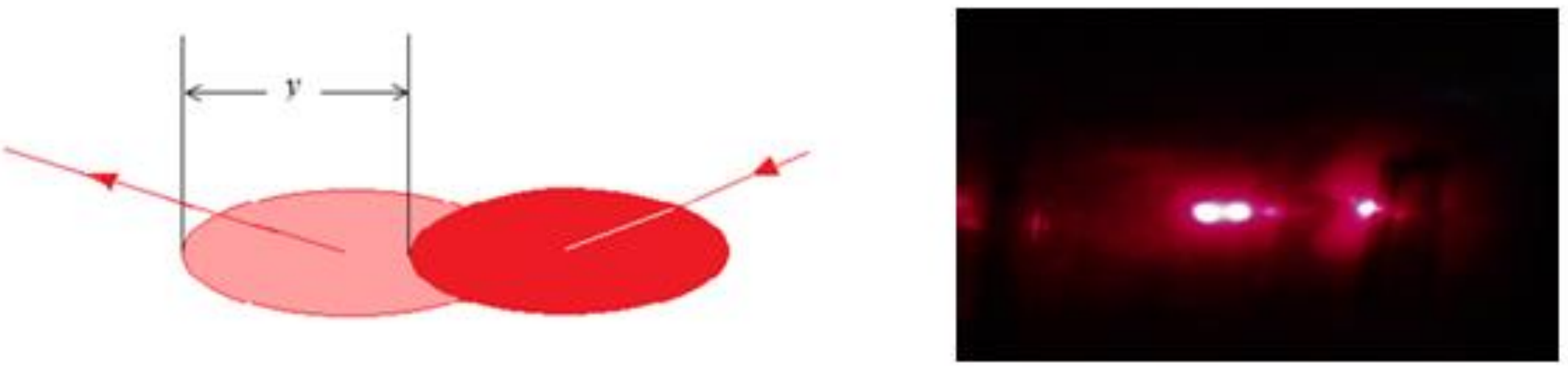

(b)

**Fig. 2**

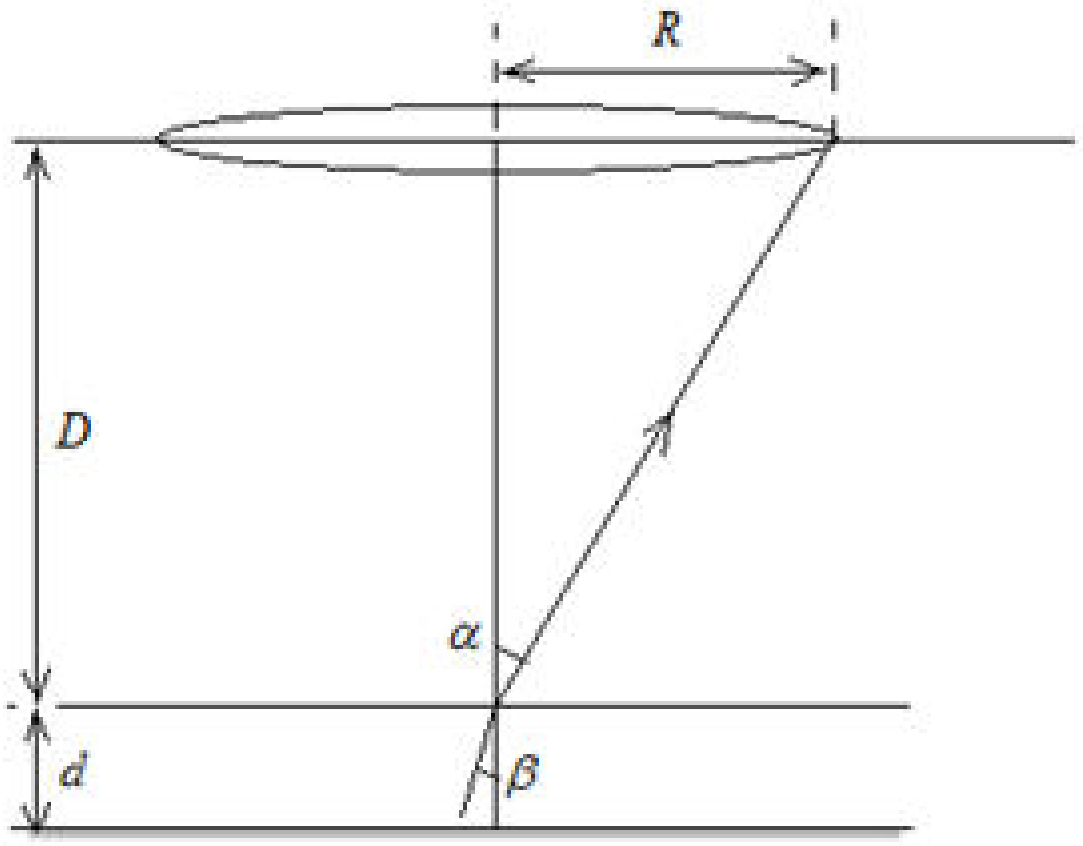

**Fig. 3**

.

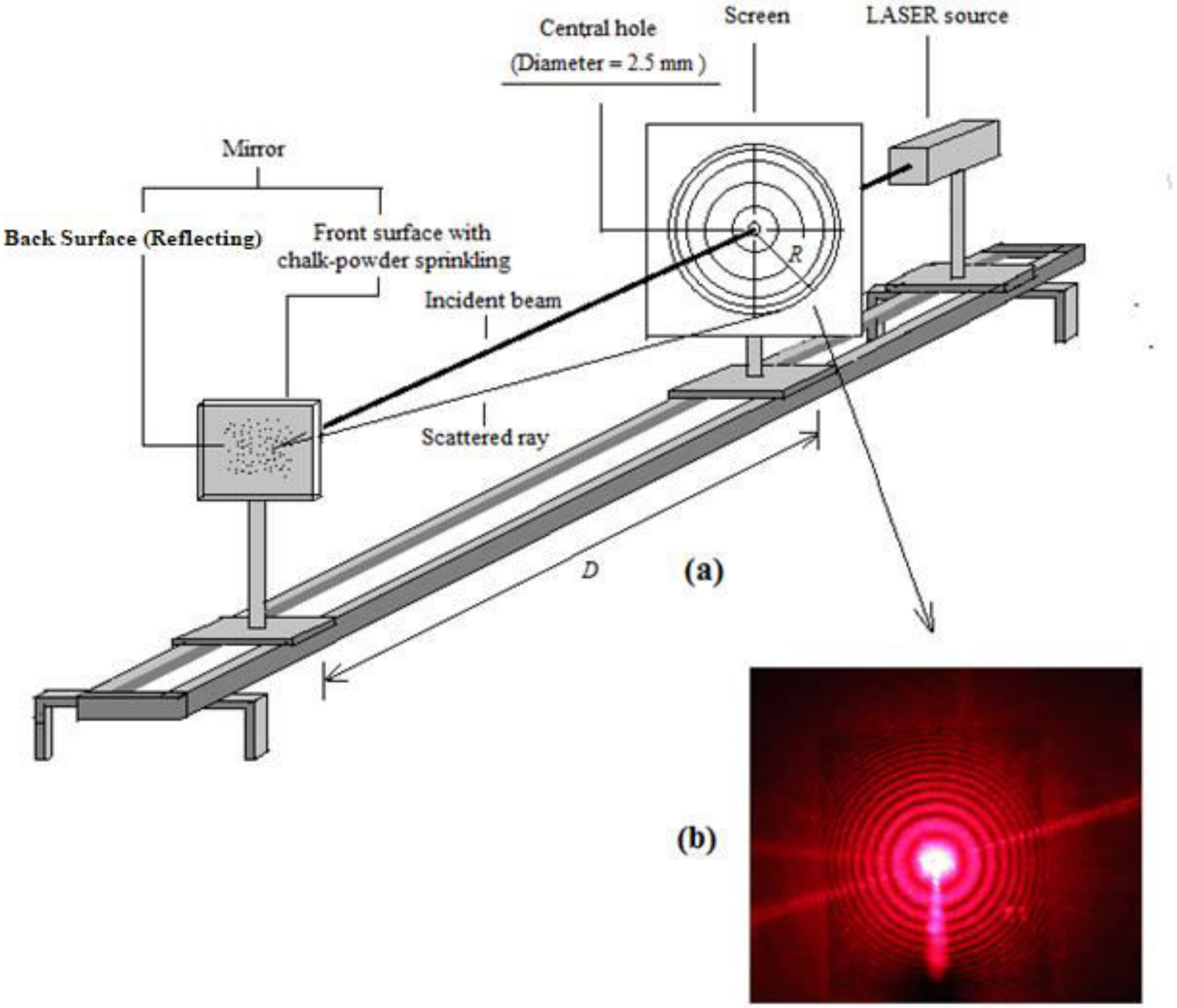

**Fig. 4**

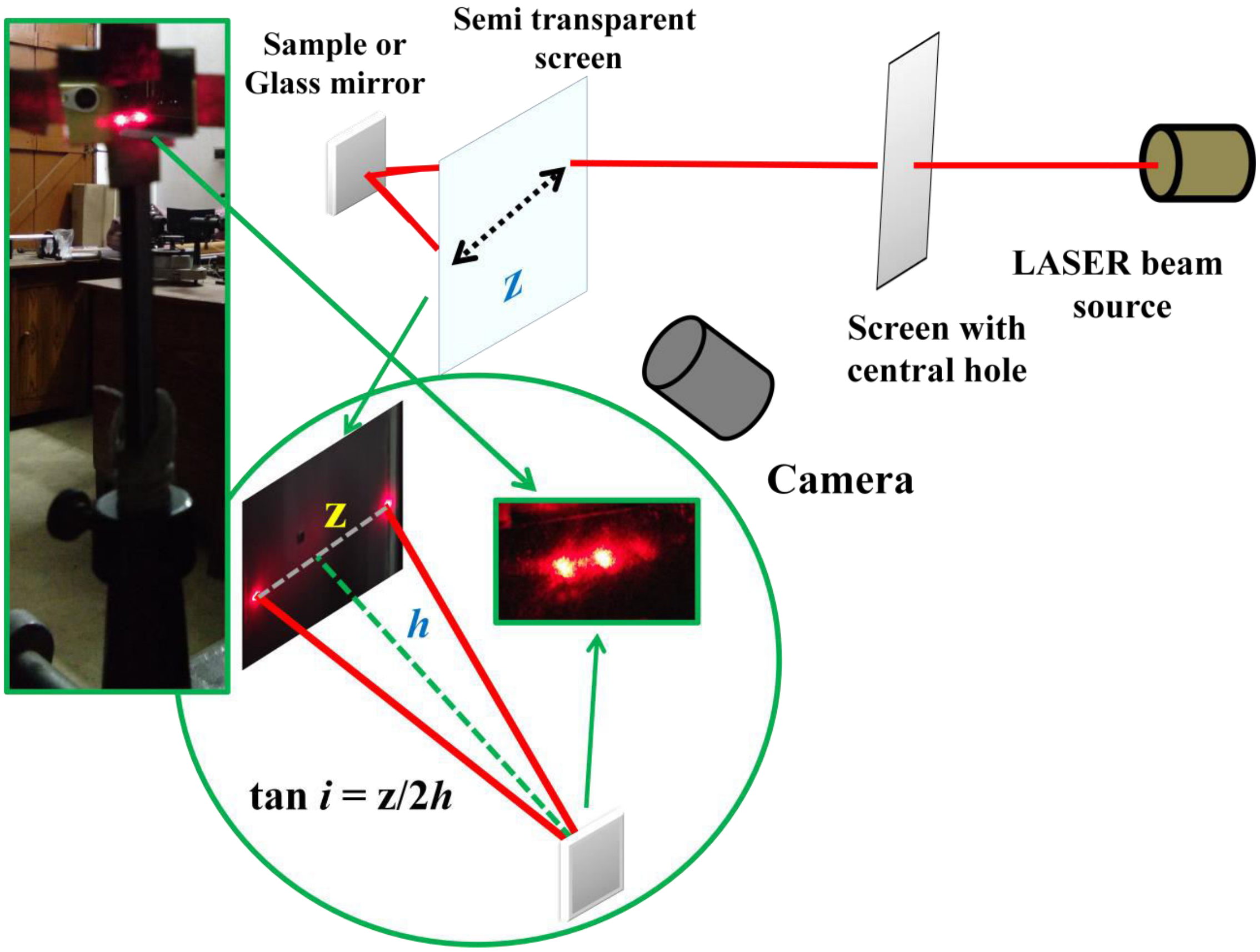

**Fig. 5**

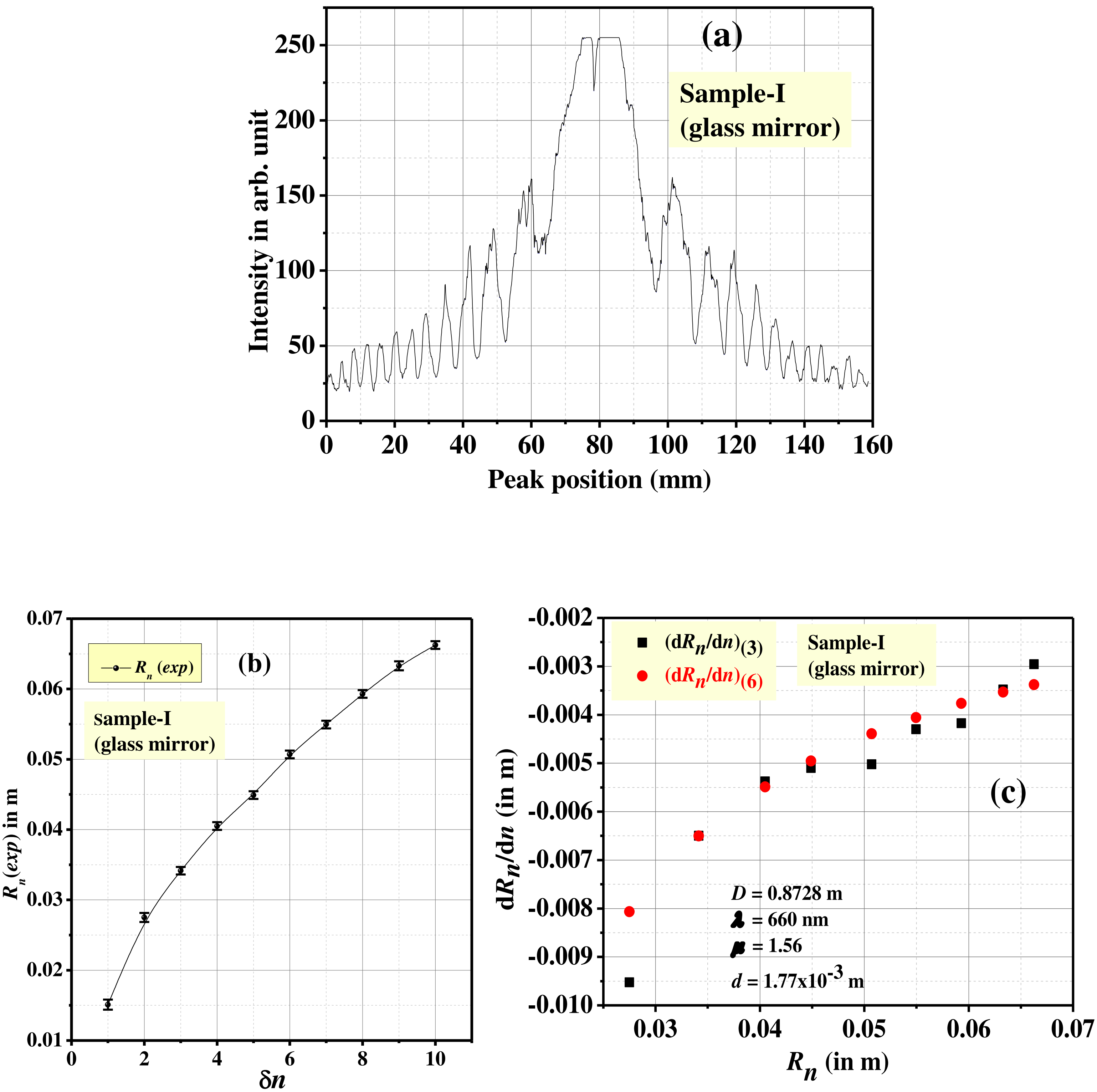

**Fig. 6**

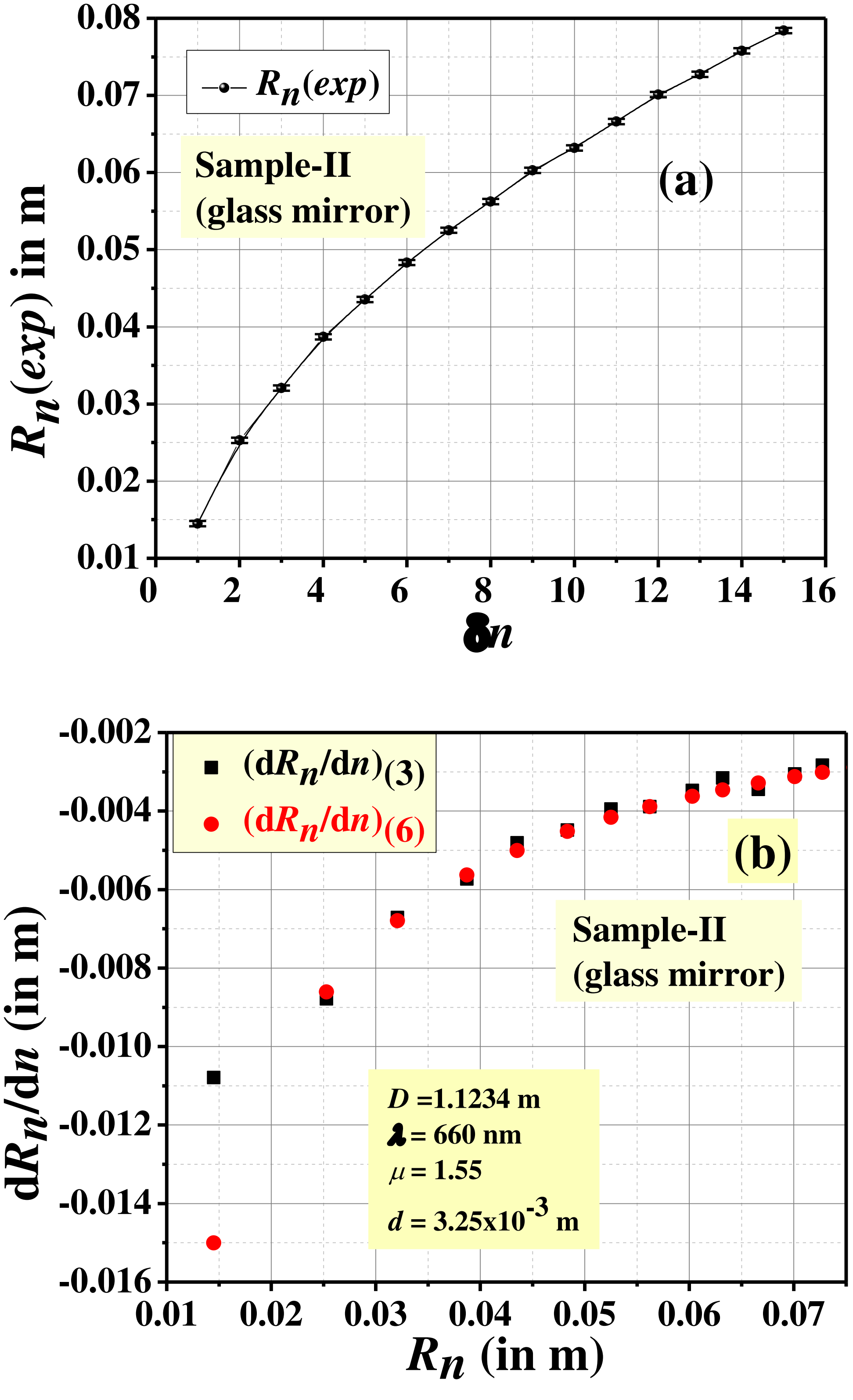

**Fig.7**

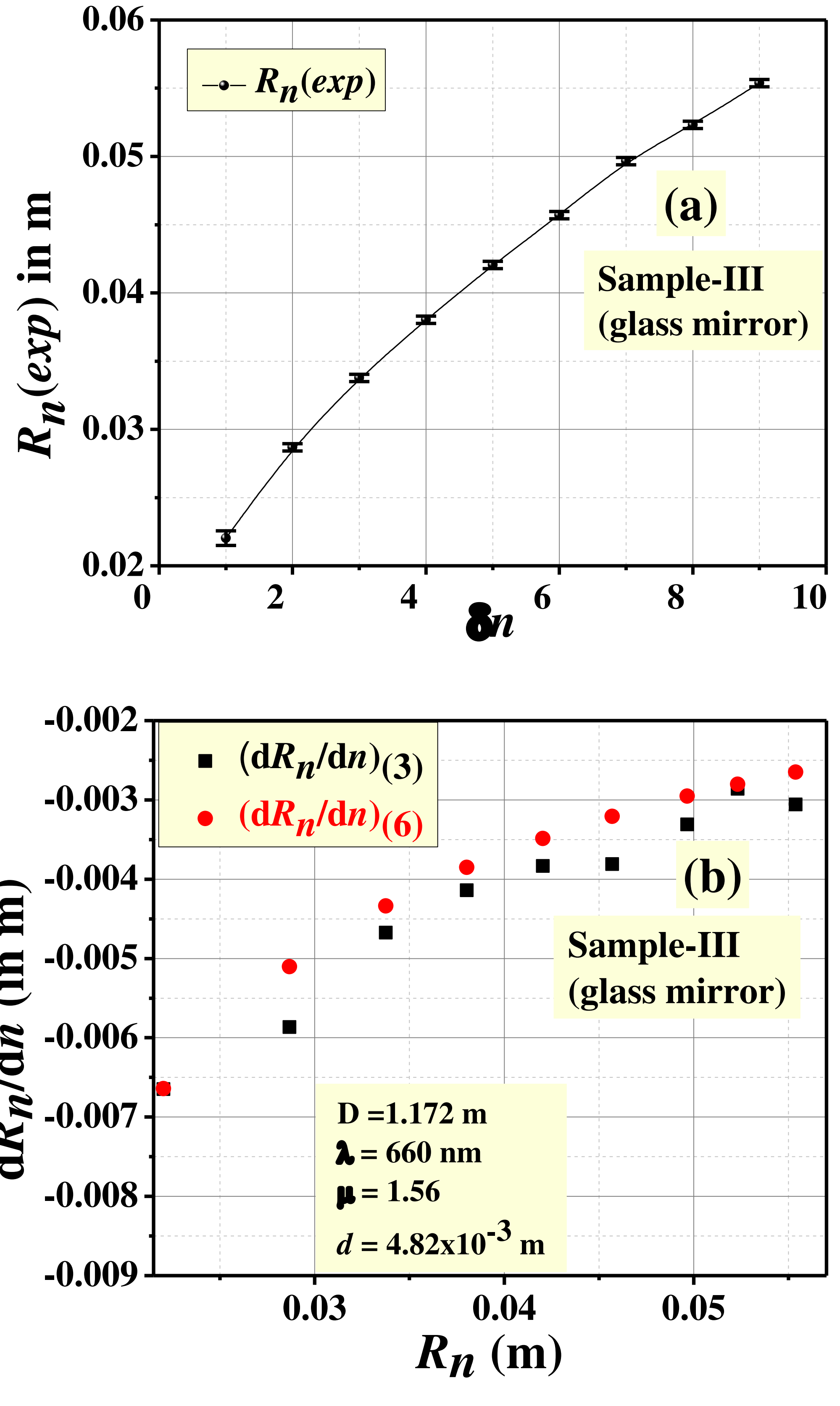

Fig. 8

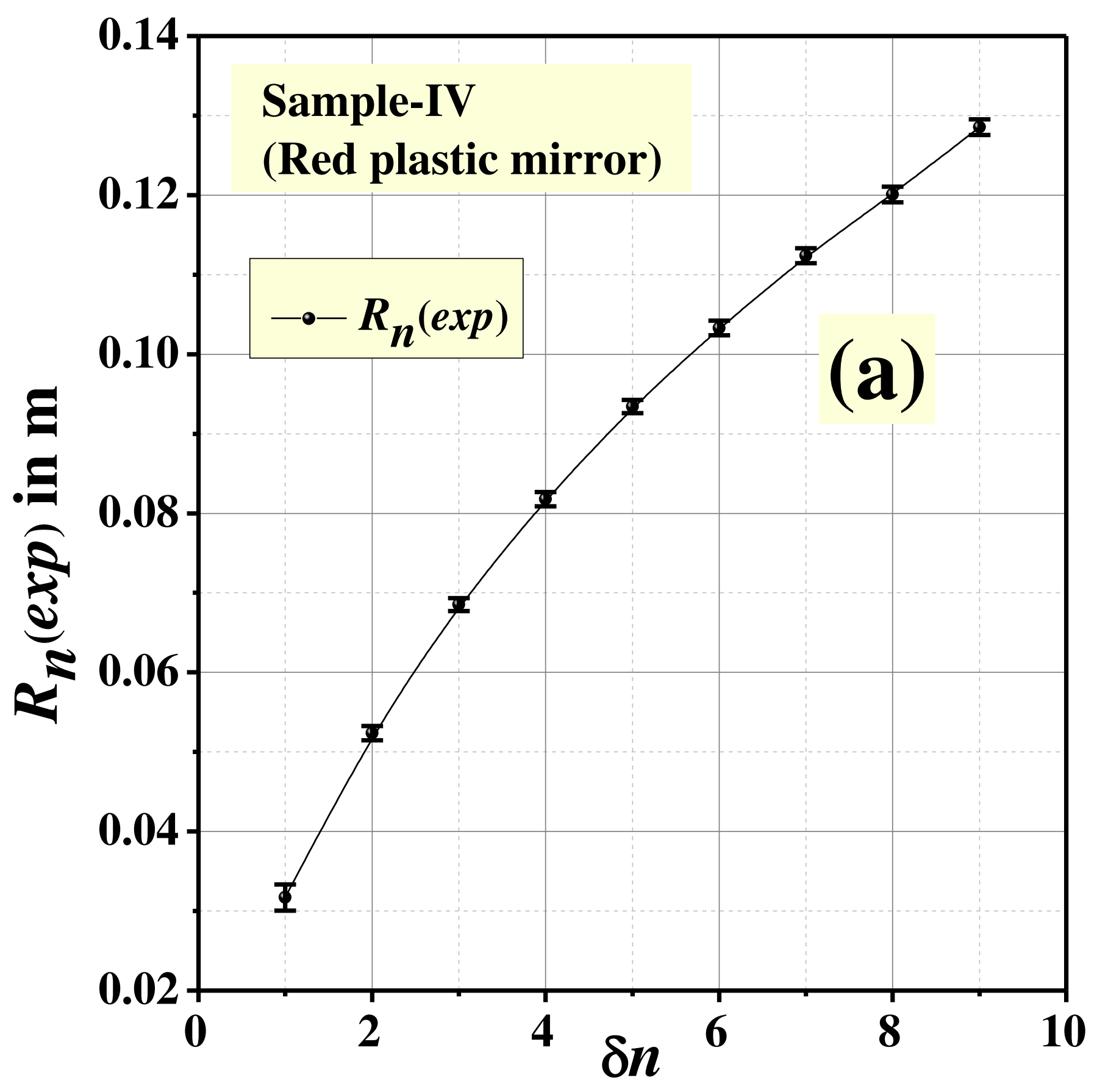

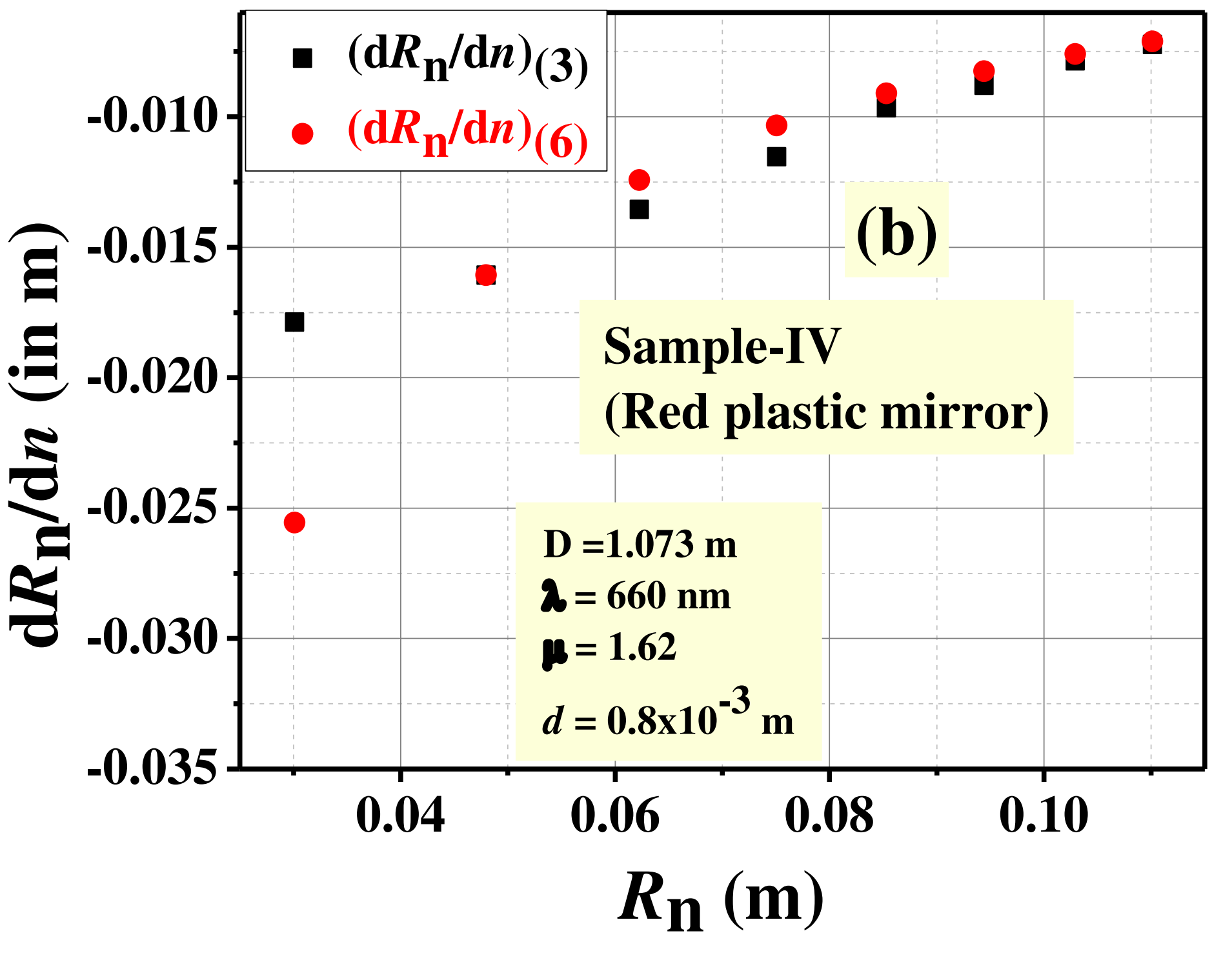

Fig.9

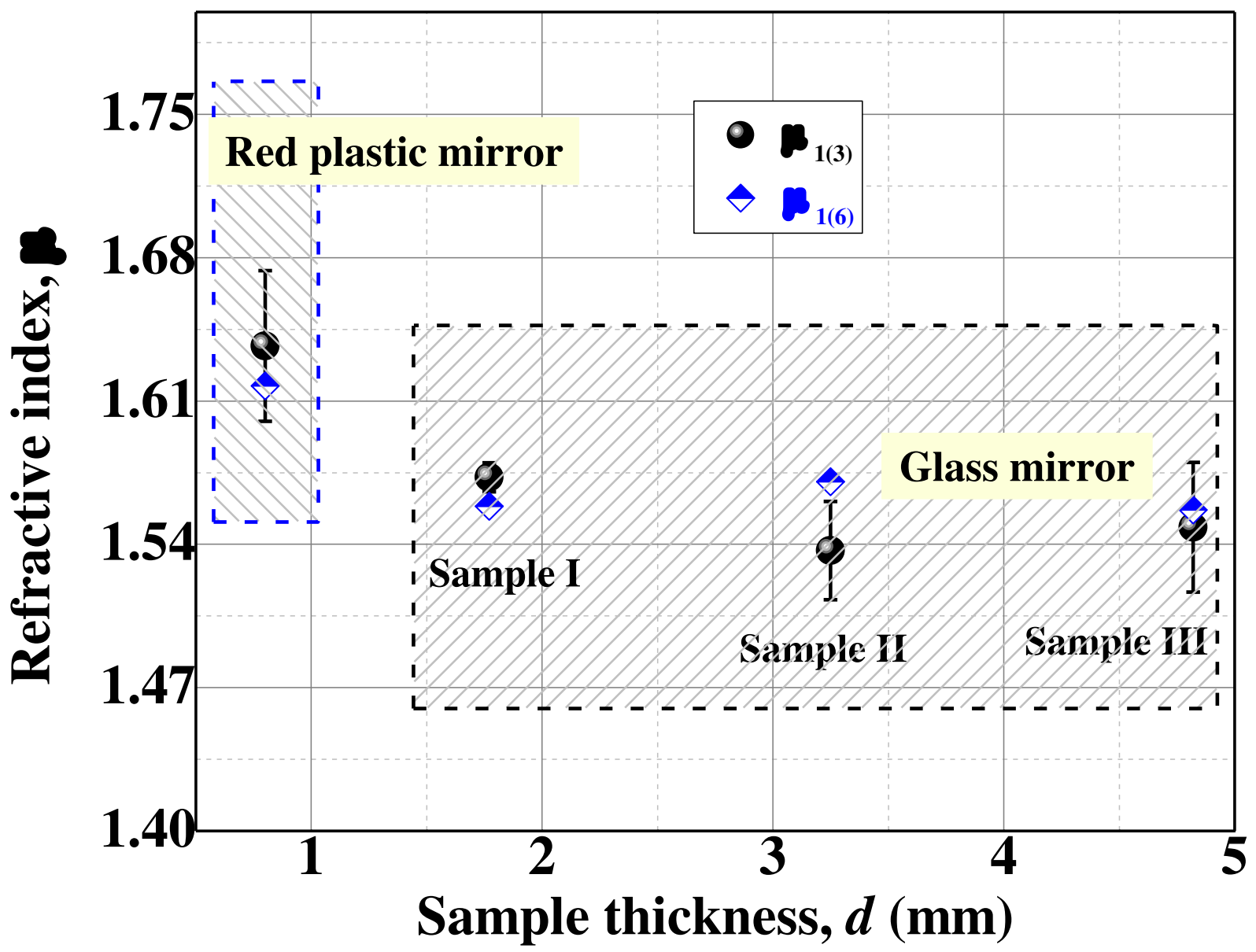

**Fig. 10**